# Pico-strain-level dynamic perturbation measurement using πFBG sensor

### DEEPA SRIVASTAVA AND BHARGAB DAS[*]

*Advanced Materials and Sensors Division, CSIR-Central Scientific Instruments Organization, Sector 30-C, Chandigarh-160030, India*
[*]*Email ID: bhargab.das@gmail.com/ bhargab.das@csio.res.in*

Interferometric interrogation technique realized for conventional fiber Bragg grating (FBG) sensors is historically known to offer the highest sensitivity measurements, however, it has not been yet explored for $\pi-$phase-shifted FBG ($\pi$FBG) sensors. This, we believe, is due to the complex nature of the reflection/transmission spectrum of a $\pi$FBG, which cannot be directly used for interferometric interrogation purpose. Therefore, we propose here an innovative as well as simple concept towards this direction, wherein, the transmission spectrum of a $\pi$FBG sensor is optically filtered using a specially designed fiber grating. The resulting filtered spectrum retains the entire characteristics of a $\pi$FBG sensor and hence the filtered spectrum can be interrogated with interferometric principles. Furthermore, due to the extremely narrow transmission notch of a $\pi$FBG sensor, a fiber interferometer can be realized with significantly longer path difference. This leads to substantially enhanced detection limit as compared to sensors based on a regular FBG of similar length. Theoretical analysis demonstrate that high resolution weak dynamic strain measurement down to $4\ p\varepsilon/\sqrt{Hz}$ is easily achievable.

***OCIS codes:*** *Fiber optic sensors; Wavelength conversion devices; Fiber Bragg gratings; $\pi-$phase-shifted fiber Bragg gratings; Interferometry*

Optical fiber based sensors have become an indispensable tool for high sensitivity measurement of both static and dynamic strain with detection limits ranging from 10$^{-12}$ to 10$^{-6}$ $\varepsilon\ Hz^{-1/2}$, where $\varepsilon$ is the fractional length change. Such low level strain measurement of relative displacement and deformations is of utmost importance in many research fields encompassing underwater acoustic array detectors, structural health monitoring of civil and aerospace edifices, ultrasonic hydrophones for medical sensing, seismic sensors for geophysical surveys etc. [1-3]. Bragg grating based fiber optic sensors are playing a significant role towards the realization of these ultrasensitive static and dynamic strain detectors. FBGs encode the physical parameter to be measured into the wavelength shift of their reflection peak known as Bragg wavelength. Sophisticate instruments are required for accurate measurement of wavelength shift especially for small amplitude dynamic perturbation sensing. The measurement resolution or the detection limit of an FBG sensor system is mostly governed by two factors: the sensitivity characteristics of the sensor head and the wavelength shift measurement resolution of the FBG peak. The sensitivity characteristics of the FBG sensor head can be improved by introducing different types of packaging techniques. Whereas, to improve the spectral measurement resolution (i.e. minimum detectable change), the FBG peak needs to be narrowed because the wavelength shift measurement resolution is highly dependent on the spectral linewidth of the grating sensor. Narrower linewidth of the Bragg reflected light leads to higher spectral resolution thereby substantially improving the minimum detectable change of the measurand. Unfortunately, the spectral linewidth of a reasonable length FBG is relatively wide with typical bandwidth (FWHM) of > 100 *pm*. This limits the spectral measurement resolution and thereby making it difficult to measure small amplitude dynamic strain measurement down to pico-strain ($p\varepsilon$) level.

A special type of FBG whose reflection spectrum features a notch caused by a $\pi$-phase discontinuity in the center of the grating (called $\pi-$phase-shifted FBG or $\pi$FBG) have attracted a great deal of attention for their enhanced detection limit leading to the measurement of weak strain perturbations [3-11]. The superior detection limit is due to the extremely narrow spectral notch (FWHM ~ 10 *pm*) of the $\pi$FBG reflection/transmission spectrum that allows high resolution measurement of spectral separation. Furthermore, the $\pi-$phase-shift region at the center of the grating decreases the effective length of the sensor making it extremely suitable for high frequency ultrasonic detection [4-6, 9, 11]. Figure 1 shows the transmission spectrum of a simulated $\pi$FBG of length 25 *mm*, which is characterized by a very narrow linewidth transmission notch at the center. For comparison, the reflection spectrum of a regular FBG of similar length is also presented. The position of the narrow spectral transmission peak formed in a $\pi$FBG is located at the Bragg wavelength ($\lambda_B$) of the grating and is governed by the grating equation,

$$\lambda_B = 2n_{eff}\Lambda \qquad (1)$$

Where, $n_{eff}$ is the effective refractive index of the optical mode propagating along the fiber and $\Lambda$ is the grating period. Dynamic strain perturbation acting on the optical fiber sensor modulates the grating period and hence can be measured by monitoring the spectral shift of the narrow transmission peak of a πFBG sensor.

The considerably narrow spectral notch (or transmission peak) of a πFBG sensor demands for special approaches for the measurement of their Bragg wavelength shift. This is because the conventional angular interrogation scheme which use a dispersive optical element and a 1D or 2D detector array does not provide enough wavelength shift measurement resolution for this type of sensors. At present, there are two mainstream methods to demodulate the πFBG Bragg wavelength shift. The basic principle of the first type of interrogation system is based on edge filter detection scheme wherein an extremely narrow linewidth (~ 0.1 $pm$) tunable laser source (TLS) is adjusted in the linear region of the reflection/transmission notch. Compared to the spectrum of the πFBG, the linewidth of the TLS is so narrow that it can be treated as a single wavelength. Strain signal encoded in shift of the πFBG spectra is demodulated by observing the reflected/transmitted optical power modulation with the help of a high speed photodetector [4, 5, 8, 9]. Another reported concept, albeit conceptually similar, is the use of two cascaded and identical πFBGs, where one πFBG is used a sensor and the other as a filter [6]. Experimental demonstrations have reported the measurement of dynamic strain down to only nano-strain ($n\varepsilon$) levels using edge filter based interrogation technique [8].

In the second method, the πFBG sensor is interrogated by a frequency/phase modulated light from a TLS using Pound-Drever-Hall (PDH) method [1, 3, 10, 11]. The technique is employed to extract the PDH error signal, which reflects the frequency detuning between the laser frequency and the π-phase-shifted Bragg grating resonance. Ultrahigh resolution measurements corresponding to dynamic strain detection with πFBG sensor down to 5 $p\varepsilon\, Hz^{-1/2}$ has been already demonstrated using PDH [3, 11]. Both these above mentioned interrogation methods require a tunable laser source in order to precisely monitor the πFBG wavelength shift which results in the disadvantage of high cost. Moreover, in order to suppress the low-frequency drift of the TLS frequency and environmental noise during scanning, a tight wavelength tracking method is also needed to lock the TLS frequency to the πFBG resonance center or to the midpoint of the linear region in the πFBG spectra. Although, PDH technique can be used to measure wavelength shifts corresponding to pico-strain ($p\varepsilon$) perturbations, the practical realization is very complicated & expensive and further does not offer multiplexing capability.

In this letter, we propose the concept of interferometric interrogation technique for high resolution measurement of spectral shift of πFBG sensors. The interferometric interrogation scheme is relatively simpler as compared to PDH architecture and simultaneously it has the ability to measure sub-pico-level ($p\varepsilon$) strain perturbations. Interferometric interrogation of regular FBG sensors is already a widely accepted technique for high resolution Bragg wavelength-shift detection [12-17]. In this case, an unbalanced fiber interferometer (e.g. Michelson or Mach-Zehnder type) is used as the wavelength-shift discriminator wherein the fiber interferometer converts the Bragg wavelength-shift of FBG sensor into a corresponding phase shift which is finally recorded as intensity variations. This fiber interferometric wavelength interrogation method has additional advantages of wide-bandwidth, high-resolution, tunable-sensitivity etc. and is more suitable for dynamic measurement of strain modulation as required in the fields of vibration and acoustics. It is noteworthy to realize that dynamic strain measurement down to nano-strain ($n\varepsilon$) level is possible to achieve using regular FBG sensors measured with interferometric interrogation technique [16, 17]. However, the broader spectral bandwidth (> 100 $pm$) of typical FBG sensors make it difficult to breach this limit.

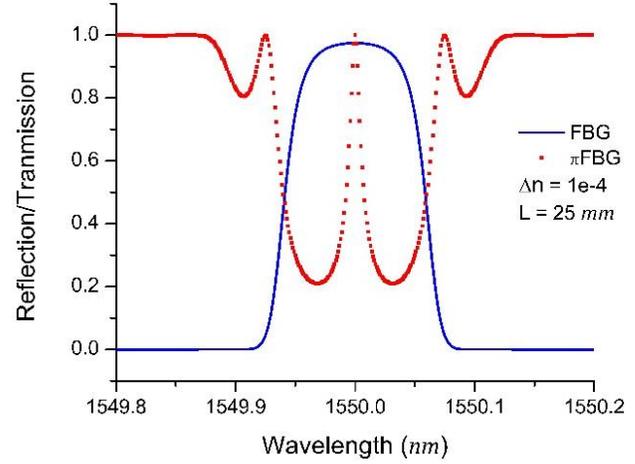

**Fig. 1: Reflection and transmission spectrum of an FBG & a πFBG.** The gratings are of similar length (L=25 $mm$), refractive index modulation ($\Delta n = 1 \times 10^{-4}$) and apodization (raised cosine profile). FWHM(FBG)~ 120 $pm$ and FWHM(πFBG)~ 11 $pm$.

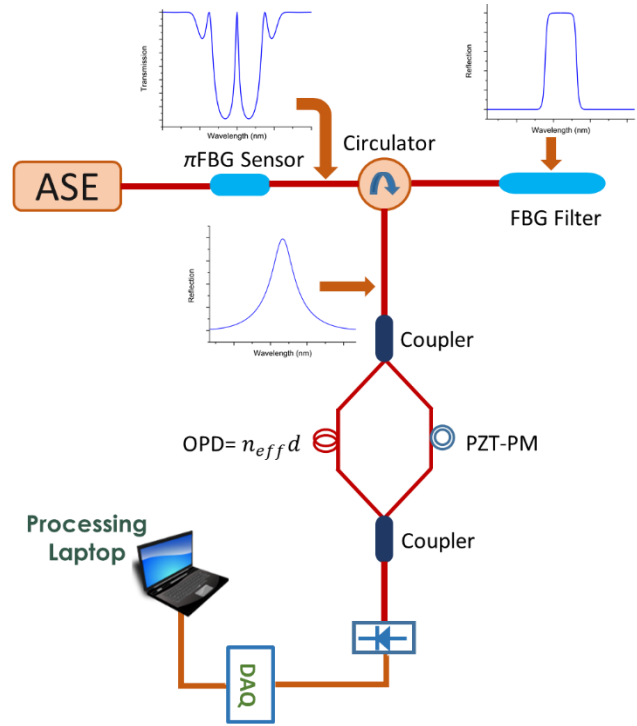

**Fig. 2: Schematic of interferometric interrogation technique for πFBG sensor.** Transmission spectrum of the πFBG sensor is directed to an FBG filter. Parameters of the FBG filter are chosen such that it reflects only the narrow spectral peak which is then guided to a fiber Mach-Zehnder interferometer. ASE: Broadband light source; OPD: Optical Path Difference; PZT-PM: Lead-Zirconate-Titanate Phase Modulator; DAQ: Data Acquisition

Card. PZT-PM is used for phase generated carrier (PGC) demodulation.

Nevertheless, the implementation of interferometric interrogation scheme for $\pi$FBG sensors is not as straight forward as for a regular FBG sensor. Whereas the reflection spectrum of FBG sensors are used for interferometric interrogation purpose, the same cannot be utilized for $\pi$FBG sensors. The reflection spectrum of $\pi$FBGs lack any peak that can be used as source input to an interferometer. Thus for a $\pi$FBG sensor, it is essential to exploit the transmission spectrum. The narrow spectral peak appearing in the central position of the total transmission spectrum of a $\pi$FBG sensor (as shown in Fig. 1) needs to be first extracted. This is proposed to be performed by optical filtering the $\pi$FBG sensor output with the help of a suitably designed FBG filter. The schematic diagram presented in Fig. 2 demonstrates this concept. The design parameters of the FBG filter *i. e.* bandwidth, center wavelength, refractive index modulation etc. are chosen in such a way that the reflected light retains only the central narrow spectral peak. This removes the undesired spectral content of $\pi$FBG transmission spectrum. A numerical model is implemented to theoretically simulate the interrogation process. The $\pi$FBG is modeled by the transfer matrix method, in which the grating is divided into a number of subsections, each associated with a 2 × 2 matrix and a phase shift matrix that accounts for the $\pi$ phase shift region of the grating [18]. The $\pi$FBG sensor is assumed to be of length 25 *mm* and the $\pi$ phase-shift region is located exactly at the center. A raised cosine apodization profile of the refractive index modulation ($\Delta n$) is assumed with peak modulation index being 1×10$^{-4}$ and $n_{eff} = 1.45$. The transmission spectrum of this $\pi$FBG sensor is shown in Fig. 1. For our interferometric interrogation purpose, we are interested only in the central part of this spectrum containing the narrow spectral peak. Figure 3 shows this narrow transmission notch which has an FWHM of ~ 11 *pm*, significantly lower than the FWHM of a conventional FBG sensor.

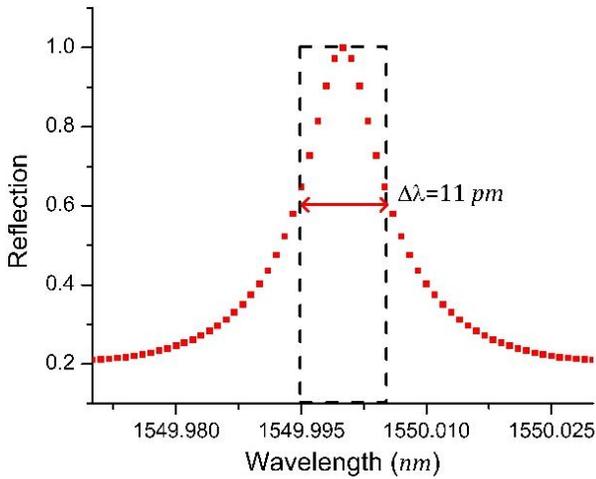

**Fig. 3: Narrow transmission notch of the $\pi$FBG sensor.** The calculated FWHM ($\Delta\lambda$) of this transmission notch is ~ 11 *pm*, which is almost an order of magnitude smaller than a regular FBG sensor reflection spectrum.

In order to isolate this narrow transmission notch, an appropriate FBG based filter is designed. The FBG filter has a length of 60 *mm* together with a raised cosine apodization profile and a peak index modulation of 0.8×10$^{-4}$. The reflection spectra of this FBG filter is shown in Fig. 4, wherein, the transmission spectra of the $\pi$FBG sensor for five different Bragg wavelengths with a spacing of 10 *pm* are also shown. This depicts the shifting of transmission notch around the central position of 1550 *nm*. The designed FBG filter is placed at one end of a circulator as shown in Fig. 2 and the reflected signal is fed to the fiber interferometer for subsequent measurement and analysis. It is important to note that the designed FBG filter has a flat-top profile which is very necessary to avoid any possible intensity variation when the transmission notch shifts under the influence of external perturbation. Figure 5 shows the spectrum of the $\pi$FBG sensor output after reflection from the designed FBG filter, which acts as a source for the Mach-Zehnder fiber interferometer for interrogation purpose. The reflection spectra are shown for wavelength shift of ±10 *pm* on either side of the central position of 1550 *nm*. As the overall objective is the measurement of very small amplitude dynamic strain perturbation, a wavelength shift of 10 *pm* of the transmission notch around the central position is sufficient enough.

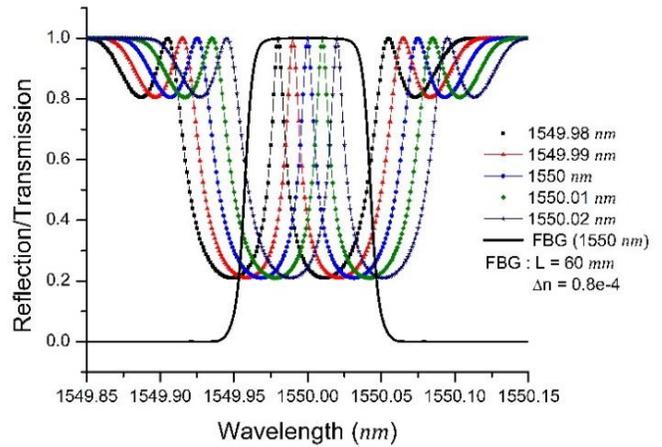

**Fig. 4: Reflection spectra of the designed FBG filter.** It has a flat top reflection profile designed to avoid intensity variation. The transmission spectra of the $\pi$FBG sensor for five different Bragg wavelengths with a spacing of 10 *pm* are also shown.

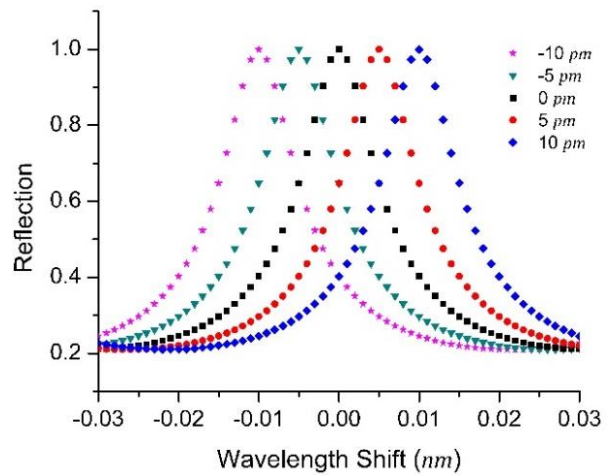

**Fig. 5: Spectrum of the $\pi$FBG sensor output after reflection from the designed FBG filter.** A wavelength shift of ±10 *pm* on either side of the central position of 1550 *nm* is depicted.

Additionally, we have calculated the integrated light intensity as the $\pi$FBG sensor transmission notch shifts around the central position. The results are shown in Fig. 6, which indicates that within the region of interest *i.e.* a wavelength shift of ±10 *pm*, the $\pi$FBG sensor transmission notch does not undergo any intensity modulation. This is very important in the case of interferometric interrogation technique as the wavelength-shift modulation is finally recorded as intensity modulation at the output of fiber interferometer. The significant reduction in the bandwidth ($\Delta\lambda$: FWHM) of the sensor light is therefore exploited to maintain a longer optical path difference (OPD) between the two arms of the fiber Mach-Zehnder interferometer (F-MZI). As the phase shift responsivity of a fiber interferometer is directly related to the OPD, an increased OPD leads to better wavelength-shift detection sensitivity. The optimum OPD for maximum sensitivity can be determined with the following relation, which was suggested by Weis *et al.* [13]:

$$OPD \times \Delta k = 2.355 \quad (2)$$

Where, $\Delta k$ is the spectral bandwidth of the optical signal fed to the fiber interferometer expressed in wave-number unit. By substituting, $\Delta\lambda = 11\ pm$ (corresponding to the transmission notch of the $\pi$FBG sensor after reflection from the FBG filter) and $\lambda = 1550\ nm$, this optimum OPD is determined to be $\sim 80\ mm$. This is significantly higher as compared to a regular FBG sensor of bandwidth $\sim 120\ pm$, which allows an optimum OPD of only 7.5 *mm*.

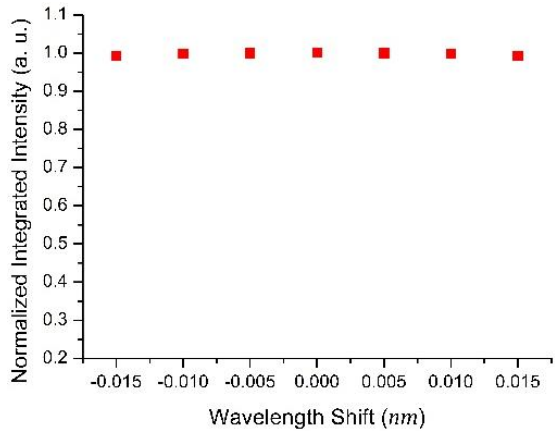

**Fig. 6: Integrated light intensity as the transmission notch of $\pi$FBG sensor shifts around the central position *i. e.* the unperturbed state.** A flat-top profile of the FBG filter within the spectral shift region is necessary for this.

For dynamic strain induced modulation in the $\pi$FBG sensor spectral notch wavelength represented by $\Delta\lambda \sin\omega t$, the phase-shift modulation [$\Delta\phi(t)$] of the fiber interferometer can be represented as:

$$\Delta\phi(t) = -\frac{2\pi\ OPD}{\lambda^2}\Delta\lambda \sin\omega t$$
$$= -\frac{2\pi\ OPD}{\lambda^2}\gamma\ \Delta\varepsilon \sin\omega t \quad (3)$$

Where, $\Delta\varepsilon$ is the dynamic strain modulation and $\gamma$ is the strain-to-wavelength shift responsivity of the $\pi$FBG grating sensor. As $\pi$FBG sensors are fabricated in the same type of optical fiber (*i.e.* silica) as used for normal FBGs, their strain-to-wavelength shift responsivity is also similar. According to the empirical model proposed by Kersey *et al.* [17], the strain responsivity is $\gamma = 1.2\ pm/\mu\varepsilon$ at 1550 *nm*. Taking this into account, together with optimum OPD of 80 *mm* and a wavelength of 1550 *nm*, would yield a sensitivity $\Delta\phi/\Delta\varepsilon$ of 0.25 $rad/\mu\varepsilon$. As typical dynamic phase shift detection of $\sim 10^{-6}\ rad/\sqrt{Hz}$ is currently possible with interferometric techniques [19-20], therefore a dynamic strain resolution of $\sim 4\ p\varepsilon/\sqrt{Hz}$ is easily achievable with a $\pi$FBG sensor interrogated with interferometric principle. Thus the detection limit for dynamic strain sensing can be significantly improved with the proposed methodology. Further, it is interesting to realize that the wavelength shift of $\pi$FBG sensor transmission notch corresponding to 4 $p\varepsilon$ is $\sim 4.8 \times 10^{-6}\ pm$. This shows the wavelength-shift measurement resolution capability of the demonstrated technique.

In conclusion, we have proposed an innovative concept of interrogating $\pi$FBG sensors using interferometric principle. External perturbation induced modulation of the narrow transmission notch is monitored by using an unbalanced fiber interferometer as a wavelength discriminator. This new technique has lesser complexity as compared to PDH method without compromising the strain measurement resolution. Leveraging the smaller bandwidth of the narrow transmission notch, which allows for a significantly longer OPD, it is theoretically established that dynamic strain resolution down to $4\ p\varepsilon/\sqrt{Hz}$ is easily achievable. Further enhancement is possible by increasing the OPD as the coherence length $\{\lambda^2/(n_{eff}\Delta\lambda)\}$ for spectral bandwidth of 11 *pm* is $\sim 150\ mm$ and the calculated optimum OPD is still far away from this value. An as alternative approach, the FBG filter can also be replaced with a low chirp rate FBG of suitable bandwidth and flat-top reflection profile. The proposed interferometric interrogation technique for π-phase-shifted Bragg grating sensor has potential applications in the field of ultra-high resolution dynamic strain measurement systems such as hydrophones for SONAR & medical sensing, acoustic emission studies etc.